\documentclass[letterpaper]{revtex4}
\usepackage[dvips]{epsfig}
\usepackage{amsmath,amssymb,bm}
\newcommand*{\vect}[1]{\bm{#1}}
\newcommand*{\unitvect}[1]{\hat{\bm{#1}}}
\begin{document}
\title{Potential of Mean Force between a Spherical Particle Suspended in
a Nematic Liquid Crystal and a Substrate}
\author{Evelina B. Kim}
\author{Roland Faller}
\author{Qiliang Yan}
\author{Nicholas L. Abbott}
\author{Juan J. \surname{de Pablo}}
\affiliation{Department of Chemical Engineering, University of Wisconsin,
Madison, WI 53706, USA}
\begin{abstract}
We consider a system where a spherical particle is suspended in a nematic 
liquid crystal confined between two walls. We calculate the liquid-crystal 
mediated potential of mean force between the sphere and a substrate by means 
of Monte Carlo simulations. Three methods are used: a traditional Monte Carlo 
approach, umbrella sampling, and a novel technique that combines canonical 
expanded ensemble simulations with a recently proposed density of states 
formalism. The latter method offers clear advantages in that it ensures good 
sampling of phase space without prior knowledge of the energy landscape of the
system. The resulting potential of mean force, computed as a function of the 
normal distance between the sphere and a surface, suggests that the sphere is 
attracted to the surface, even in the absence of attractive molecular
interactions.
\end{abstract}
\maketitle
\section{Introduction}
The alignment of nematic liquid crystals can be controlled by surface
chemistry or topography on the scale of a few nanometers. When the uniform 
alignment is disturbed, this event can be detected through the appearance of
multidomains in an optical microscope between cross-polarizers. One possible
event that can trigger this transition is the binding of a protein or a virus 
to the surface, thereby providing the basis for development of
sensors~\cite{gupta98,vannelson02}. Recent experiments have in fact shown that
such liquid crystal sensors are remarkably sensitive to the presence of minute
amounts of protein on a substrate~\cite{gupta98,vannelson02}.

Several experiments using colloids in bulk or confined liquid crystals have
studied the influence of local disturbances in a nematic
host~\cite{poulin97,poulin98,gu00}. Colloidal particles in liquid crystals
have also been studied by computer simulations~\cite{billeter00,andrienko01}. 
One of the questions that arises from these experiments pertains to the nature
of the liquid crystal mediated interaction between the colloids, or a 
colloidal particle and a surface.

This interaction can be quantified through a potential of mean force; molecular
simulations provide a means to calculate it directly by fixing or constraining
a reaction coordinate of interest. In the particular case of liquid crystals, 
however, conventional simulation techniques such as Monte Carlo (MC) or 
molecular dynamics (MD) in regular ensembles are limited in their ability to
sample phase space. The system can be trapped in local energy minima and, for 
MD simulations, the time scales associated with the motion of the colloid can 
be prohibitively long.

The majority of available methods for measuring the potential of mean force 
either require prior knowledge of the free energy of a
system~\cite{roux94,engkvist96,vondele00}, or constrain the system to a
limited range of the reaction coordinate~\cite{straatsma92,sprik98}.
A novel method is proposed in this work to calculate the potential of
mean force which does not require prior knowledge of the free energy
and permits efficient sampling. The proposed technique consists of an expanded
ensemble in the reaction coordinate, implemented within the formalism
of a recently proposed density-of-states sampling
technique~\cite{wang01a}. Related methods have been proposed by de Oliveira
{\it et al.}~\cite{deoliveira96} and by 
Engkvist {\it et al.}~\cite{engkvist96}. This method is applied here to
determine the potential of mean force between a spherical colloidal particle
suspended in a confined nematic liquid crystal and a wall.
\section{Simulation Methods}
\subsection{Potential of Mean Force}
The potential of mean force (PMF) provides a measure of the effective
difference in free energy between two states as a function of one or several 
``interesting'' degrees of freedom. These degrees of freedom can be real 
coordinates of the system or a combination of them. They are often called 
reaction coordinate(s) $\xi$. An integration of the derivative of the 
Hamiltonian with respect to the coordinate of interest is performed along the 
reaction coordinate to provide a local estimate of the PMF, $w$, between two 
points $\xi_1$ and $\xi_2$:
\begin{equation}
 w(\xi_2)-w(\xi_1)=\int_{\xi_1}^{\xi_2}d\xi^{\prime}
 \biggl\langle\frac{\partial H}{\partial \xi}\biggr\rangle
\end{equation}
The reaction coordinate $\xi$ need not be directly observable, in the case of 
chemical reactions, for example, it can be the extent of reaction, and the 
potential of mean force provides a measure of the activation energy. It can 
also be a physical coordinate, such as the distance between two solute 
molecules in a solvent---the PMF then reflects the work required to bring two 
solute particles from an infinite to a finite separation. Both examples 
involve significant energy barriers which would severely limit the 
effectiveness of conventional MD or MC simulations for its calculation.

The PMF is connected to $p(\xi)$ (the probability of finding the system of 
interest in a state corresponding to a particular value of $\xi$) through the
reversible work theorem by~\cite{chandler87}
\begin{equation}
  w(\xi)=-kT\ln p(\xi).
\end{equation}
Since the probability of visiting high energy states is low, their
sampling and consequently the estimate of an energy barrier is
poor. In order to force a system to sample the desired portion of
phase space, it is often advantageous to fix the reaction
coordinate or to apply a biasing potential that changes the free
energy in a way that samples the barrier more efficiently. The
former approach has the shortcoming of requiring a large number of
simulations, so as to obtain a reasonable resolution for subsequent
thermodynamic integration. We concentrate on the latter approach,
as it offers the possibility of considerable improvements in
sampling efficiency and accuracy.
\subsection{Expanded Density of States Method (EDOS)}
The idea of applying a biasing potential appears in many forms and
names in the literature. Umbrella sampling~\cite{torrie74,beutler94} or 
multicanonical sampling are examples of such a strategy~\cite{berg92a}. In 
these techniques, the potential of the system is artificially altered in a way
to lower free energy barriers and permit more uniform sampling of the altered 
phase space. Histograms of thermodynamic quantities are subsequently 
reweighted to recover results corresponding to the original 
potential~\cite{ferrenberg88}. Another way of traversing phase space more 
efficiently is through the use of expanded ensembles~\cite{lyubartsev92} in 
which intermediate states are introduced. These intermediate states do not 
necessarily represent states of the physical system; their role is to
facilitate transitions between states separated by large energy 
barriers~\cite{lyubartsev92,escobedo95b}. Expanded ensemble 
formalisms have the disadvantage that visits to a specific state are dictated 
through unknown weights which must be determined iteratively.

In recent years, new methods have been proposed to calculate the density of 
states of a system directly~\cite{engkvist96,deoliveira96,wang01a}. Here we 
refer to these techniques as density of states Monte Carlo methods (DSMC). 
These methods introduce a biasing potential, which is a running best estimate 
of the inverse of the density of states as a function of energy, $g(E)$. It is 
obtained by performing a random walk in energy space. The main attractive
feature of these methods is that $g(E)$ need not be known a priori; it 
converges self-consistently to the correct density of states. DSMC has been 
successfully applied to such diverse systems as Potts and Ising 
models~\cite{wang01b}, Lennard-Jones systems~\cite{yan02}, binary 
Lennard-Jones glasses~\cite{faller01sc}, proteins~\cite{rathore02}, 
polymers~\cite{jain02}, and bulk liquid crystals~\cite{kim02sb}.

In this work, the DSMC formalism is combined with an expanded ensemble where
the expanded states correspond to a reaction coordinate. In order to span the
entire range of expanded states and visit them with uniform probability during
the simulation, the states are weighted accordingly. In a sense, the weighting
factors act as a biasing potential. An idea similar in spirit was proposed 
earlier by Engkvist and Karlstr{\"o}m~\cite{engkvist96}; their method, however,
requires a specific adaptation of the way in which the weights are updated for
any new system to be investigated. Our new technique does not suffer from this
problem and therefore can be viewed as self-consistent.

Consider a system of $N$ particles in volume $V$ and at temperature $T$. The 
system is expanded in $M$ subsystems along a reaction coordinate $\xi$ such 
that the $m^{th}$ state corresponds to a particular value of $\xi=\xi_m$. For
concreteness, the reaction coordinate is set to be a distance $X$ between two 
solute particles in a solvent. A set of $M$ states is defined by assigning to 
each of them a value of $X$ in the range $[X_{m-1},X_m]$, with $X_0$ and 
$X_{m=M}$ being the absolute bounds of the region of interest. The partition 
function $\Omega$ of the expanded ensemble is given by
\begin{equation}
  \Omega=\sum _{m=1}^M Q(N,V,T,m)g_m=\sum _{m=1}^M Q_mg_m,
\end{equation}
where $Q_m$ and $g_m$ denote a canonical partition function and a
positive weighting factor, respectively, each corresponding to a
particular state $m$. During the course of the simulation solvent
and solute particles are rearranged through arbitrary Monte Carlo moves, e.g.,
trial random displacements. Whenever one of the solute particles is
displaced, the distance $X$ between two solutes assumes a certain
value that corresponds to some state $m$. The probability $p_m$
with which a state $m$ is visited is related to the partition
function through
\begin{equation}
  p_m=\frac{g_mQ_m}{\Omega}\label{eq:prop-part}.
\end{equation}
In general, for any two states $m$ and $k$, the following relationship holds:
\begin{equation}
  \frac{Q_m}{Q_k}=\frac{p_mg_k}{p_kg_m}.
\end{equation}
The free energy difference between states $m$ and $k$ can be inferred from the
weighting factors and the population of the respective states according to
\begin{equation}
  \beta[w(\xi_m)-w(\xi_k)]=-\ln\frac{Q_m}{Q_k}=-[\ln g_k-\ln g_m]+
  [\ln p_k-\ln p_m].
\end{equation}
The relative population of the states is only used as an indicator of
uniform sampling. The physical meaning of the weighting factors is
apparent in Eq.~(\ref{eq:prop-part}): if the states are sampled
uniformly, the free energy difference between states $m$ and $k$
is given by the natural logarithm of the ratio of the
corresponding weighting factors. This stresses the importance of
finding an efficient scheme that allows the weights to
converge to the their ``true'' free energy value rapidly and
accurately.

Continuing with the example of two solute particles in a solvent,
a transition from an {\it old} to a {\it new} state occurs when
the distance $X$ between the two solutes changes accordingly
(molecular rearrangement of the solvent particles does not change
the system's state). The criteria for accepting a trial move
from an {\it old} to a {\it new} state are given by
\begin{equation}
  P_{\text{acc}}(\mathit{old} \to \mathit{new})=\min
  \{1,\exp[-\beta(U_{\text{new}}-U_{\text{old}})
  -(\ln g_{\text{new}}-\ln g_{\text{old}})]\}=\min
  \{1,\exp[-\beta\Delta U-\Delta\ln g]\}\label{eq:pacc}.
\end{equation}
Every time a state is visited, the corresponding weight is modified by a 
convergence factor, $f$:
\begin{equation}
  g \to f\cdot g \label{eq:upd}
\end{equation}
and the histogram of that state is updated. Note that in practice it is more 
convenient to work with $\ln g$, rather than $g$ itself since the former is 
used directly in the acceptance criteria; hence $\ln f$ is added to
$\ln g$. According to Eqs.~(\ref{eq:pacc}) and~(\ref{eq:upd}), if
the original state has a higher energy than the new one, the new
state will be visited more easily because  $\exp(-\beta[U_{new}-U_{old}])$ is
larger than unity and $g_{\text{new}}$ will be updated more frequently than
$g_{\text{old}}$.  Eventually, $g_{\text{new}}/g_{\text{old}}$
becomes smaller than unity and forces visits to the old, higher
energy (and less visited) state. By construction, if $g$ is
underestimated, i.e. if the contribution of the state to the total
partition function is too small, the system can readily visit the
under-visited state, and its weight is increased
accordingly until a balance is reached; the converse also holds.
Once a global histogram is sufficiently flat (the minimum
population is enforced to be at least 85\% of the average), a cycle is assumed
to be ``complete'' and the convergence factor is altered according to an
arbitrary, monotonically decreasing function; in the present
implementation we use the square root. The random walk cycle is
then resumed. The flatness condition stems from the fact that, if
the inverse density of states were used as a weighting function,
the histogram of energy would be perfectly flat, as the
probability of visiting a state would be uniform.

Several issues are worth pointing out. First, in previous
applications of DSMC, simulations proceeded until $\ln f$ reached
a low threshold value; $\ln f <10^{-8}$ has been commonly used.
Such precision was important as $g$ reflected the density of
states of the entire system over many orders of magnitude in
energy. In the application discussed here, such precision is not
required; the range of free energy or $\ln g$ is much narrower (temperature is
fixed), and only a single or several reaction
coordinates are of interest. Second, as can be inferred from the
description of the method, the weights involved in the acceptance criteria are
updated continuously, and the method does not obey detailed
balance. Once the weights have been determined,
however, the simulation can be rerun in a regular expanded
ensemble without updating the weights, to verify that state
population is indeed uniform. Third, the weight updating scheme
proposed here may be used as an efficient tool for adequate
sampling of any system with inherent, high energy barriers.
If one is interested in a particular quantity that can
be measured directly during the simulation (e.g. the mean force), it is
sufficient to stop the simulation once good sampling of that
quantity is achieved. Here we are interested in the potential of
mean force; once a good estimate is obtained, the mean force can
be integrated to yield the potential of mean force.
\subsection{Umbrella Sampling}
In order to compare the results of the proposed method to more
traditional techniques, umbrella sampling simulations were
conducted over a sub-range of the reaction coordinate of interest.
The particular implementation of the umbrella potential method
employed here is discussed in what follows. As before, an
artificial potential is introduced to prevent the system from
being trapped in local energy minima. If multiple energy minima
are present, i.e. the system is characterized by a rough energy
profile, it is more effective to subdivide the entire range of the
reaction coordinate into smaller overlapping ranges or
``windows'', and to perform multiple simulations for each window.
If the reaction coordinate is a distance, as in our case, a
natural choice for the potential is a harmonic spring potential.
The rapid increase at some distance from the spring equilibrium
position guarantees that the system is contained within a chosen
distance window. Alternatively, instead of applying a
pre-determined umbrella potential, one can iteratively arrive at
an umbrella potential that should be close to $-w(\xi)$ through a
multicanonical simulation. To this end one would record the
histogram using a given umbrella potential, Boltzmann invert it to
get an estimate of the relative free energy, and add this estimate
to the potential~\cite{faller02c}.

For any given window, the PMF is defined as
\begin{equation}
\beta w (\xi)=-\ln \Big\{\frac{\int\dotsi\int e^{-\beta U}
dr_{n+1}\dotsm dr_N }{\int\dotsi\int e^{-\beta U}dr_{1}\dotsm
dr_N}\Big\}=-\ln\Big\{ \frac{Z(\xi)}{Z}\Big\}
\end{equation}
where $\xi$ is a $n$-dimensional set of reaction coordinates
related to coordinates $r_1 \dots r_n$.  If the umbrella potential
$U_{\text{umb}}$ is introduced then the PMF can be computed from:
\begin{equation}
 \beta w (\xi)=-\ln Z_{\text{umb}}(\xi)-\beta U_{\text{umb}}(\xi)-\ln Z
\end{equation}
Here $Z_{\text{umb}}$ and $Z$ correspond to the partition
functions of the system with and without an umbrella potential,
respectively. Another way to express the PMF is
\begin{equation}
  \beta w(\xi) = -\ln g_{\text{umb}}(\xi)-\beta
  U_{\text{umb}}-\ln\frac{Z_{\text{umb}}}{Z},
\end{equation}
where $g_{\text{umb}}(\xi)$ denotes the correlation function or
the probability of finding a system at a particular value of
$\xi$. Note that the last term in the above equation is a constant
(which is different for every window). Since we are interested in
the relative free energy and it is a continuous function of $\xi$,
the overlapping ranges of the curves $w(\xi)$ have to be collapsed
onto one master curve by fitting an offset for each window. Once
again, the disadvantages of the method in this rendering include
uncertainty in identifying an optimal umbrella potential, in choosing the 
number of necessary windows, and in fitting the final results for complex
systems.
\section{Simulation Details}
\begin{figure}
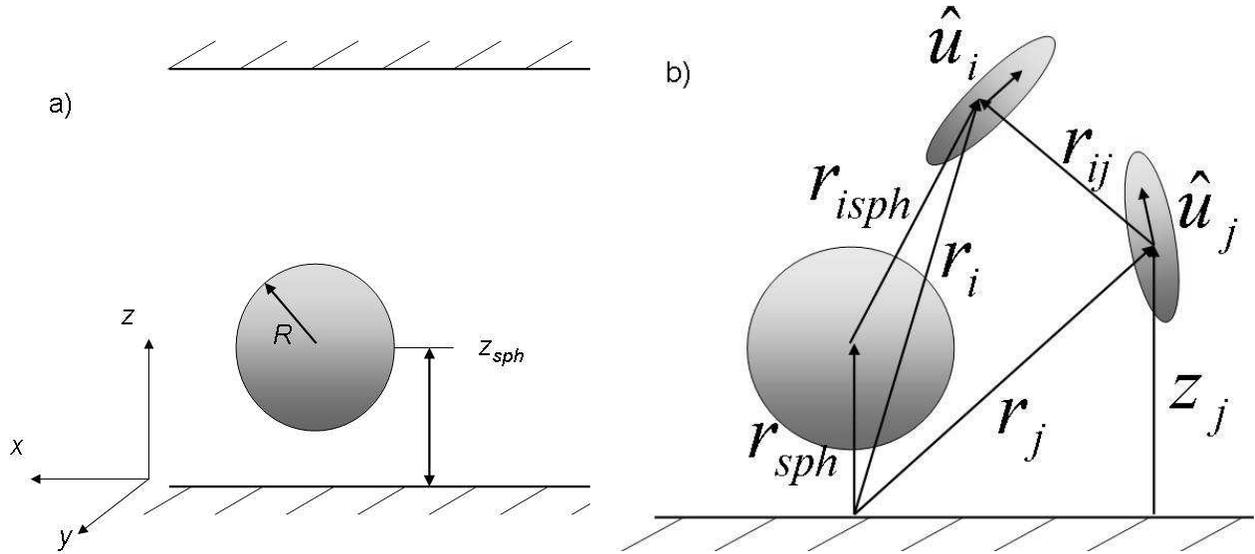

  \includegraphics[width=0.45\linewidth]{schematic.epsi}
  \includegraphics[width=0.45\linewidth]{GB_schematic.epsi}	
  \caption{ a) Schematic view of the system under consideration. b) 
	The different interactions between the mesogens and the colloid.}
  \label{fig:syssketch}
\end{figure}
The system considered in this work comprises 11460 liquid crystal particles
confined between two soft repulsive walls at $z=0$ and
$z=Z_{\text{wall}}$ (see Figure~\ref{fig:syssketch}a). Simulations were 
conducted at a constant
temperature $T^*=1.0$ and an average number density of
$\rho^*=0.335$. All variables are given in reduced units, e.g.
$\rho^* = N\sigma_0^3/V$, $T^* =kT/\epsilon_0$, where $\sigma_0$
and $\epsilon_0$ are length and energy parameters which are set to
unity. The shape of the simulation box is orthorhombic; the side
lengths are equal in the $x$ and $y$ directions. Periodic boundary
conditions apply in $x$ and $y$. Mesogens $i$ and
$j$ interact via a shifted and truncated Gay--Berne (GB)
potential~\cite{gay81}:
\begin{eqnarray}
  U_{ij}&=&\begin{cases}4\epsilon_0(\varrho_{ij}^{-12}-\varrho_{ij}^{-6})+
  \epsilon_0 ,\quad &\varrho_{ij}^6<2\\
 0,\quad &\varrho_{ij}^6>2\end{cases}\label{Eqn:mainGB}\\
  \varrho_{ij} &=& (|\vect{r}_{ij}|-\sigma_{ij}+\sigma_0)/\sigma_0
	\label{Eqn:GB_rho}\\
  \sigma_{ij} &=& \sigma_o \left[ 1 -\frac {\chi}{2} \left\{
  \frac {\left(\unitvect{r}_{ij} \cdot \unitvect{u}_i + \unitvect{r}_{ij} \cdot
\unitvect{u}_j \right)^2} { 1 +  \chi \unitvect{u}_i \unitvect{u}_j } +
  \frac {\left(\unitvect{r}_{ij} \cdot \unitvect{u}_i + \unitvect{r}_{ij} 
	\cdot \unitvect{u}_j \right)^2} { 1
  - \chi \unitvect{u}_i \unitvect{u}_j } \right\} \right] ^{-1/2}
 \label{Eqn:GB sigma}\\
 \label{Eqn:GB chi}
  \chi &=& \frac {\kappa^2 - 1} {\kappa^2 + 1}.
\end{eqnarray}
where $\unitvect{u}_i$ and $\vect{r}_{ij}$ denote molecular
orientations (along the main axis of the ellipsoid) and intermolecular
vectors ($\vect{r}_{ij} = \vect{r}_{i}-\vect{r}_{j}$, where
$\vect{r}_{i}$ is the position of particle $i$); unit vectors are
identified by hats
($\unitvect{r}_{ij}=\vect{r}_{ij}/|\vect{r}_{ij}|$). For details see
Figure~\ref{fig:syssketch}b. Parameter $\kappa=3$ in this work is the
length to width ($\sigma_0$) ratio. The interactions of a GB mesogen
$i$ with all interfaces are described by the same potential
(\ref{Eqn:mainGB}), where the $\varrho_{ij}$ are defined as follows
for a sphere and a surface:
\begin{eqnarray}
  \varrho_{i\text{sph}} &=&
  \left(|\vect{r}_i-\vect{r}_{\text{sph}}|-{R}+\sigma_0/2\right)/\sigma_0,
  \label{Eqn:sph rho}\\
  \varrho_{i\text{wall}} &=&
  \left(|\vect{z}_i|+\sigma_o/2\right)/\sigma_0.\label{Eqn:surf rho}
\end{eqnarray}
In this case, $\vect{r}_{\text{sph}}$, $\vect{r}_i$, and
$|\vect{z}_i|$ denote the position vectors for the centers of mass of
the sphere, mesogen $i$, and the normal distance from a surface,
respectively; $R$ represents the radius of the sphere.  This
potential has been employed in a previous study of a colloidal
particle in a bulk liquid crystal~\cite{andrienko01}, and is known
to reproduce topological defects around the particle that are in
agreement with theory and experiment. The force on a sphere is
calculated as the derivative of the interaction potential between
the liquid crystal and the sphere with respect to $z$; due to
symmetry in the $(x,y)$-plane, only the $z$-component of the force
contributes. The surface and sphere interact as hard bodies, i.e.
the minimum distance possible between a surface and the center of
mass of the sphere is the radius of the sphere, which is set to 3.

\begin{figure}
  \includegraphics[width=0.7\linewidth]{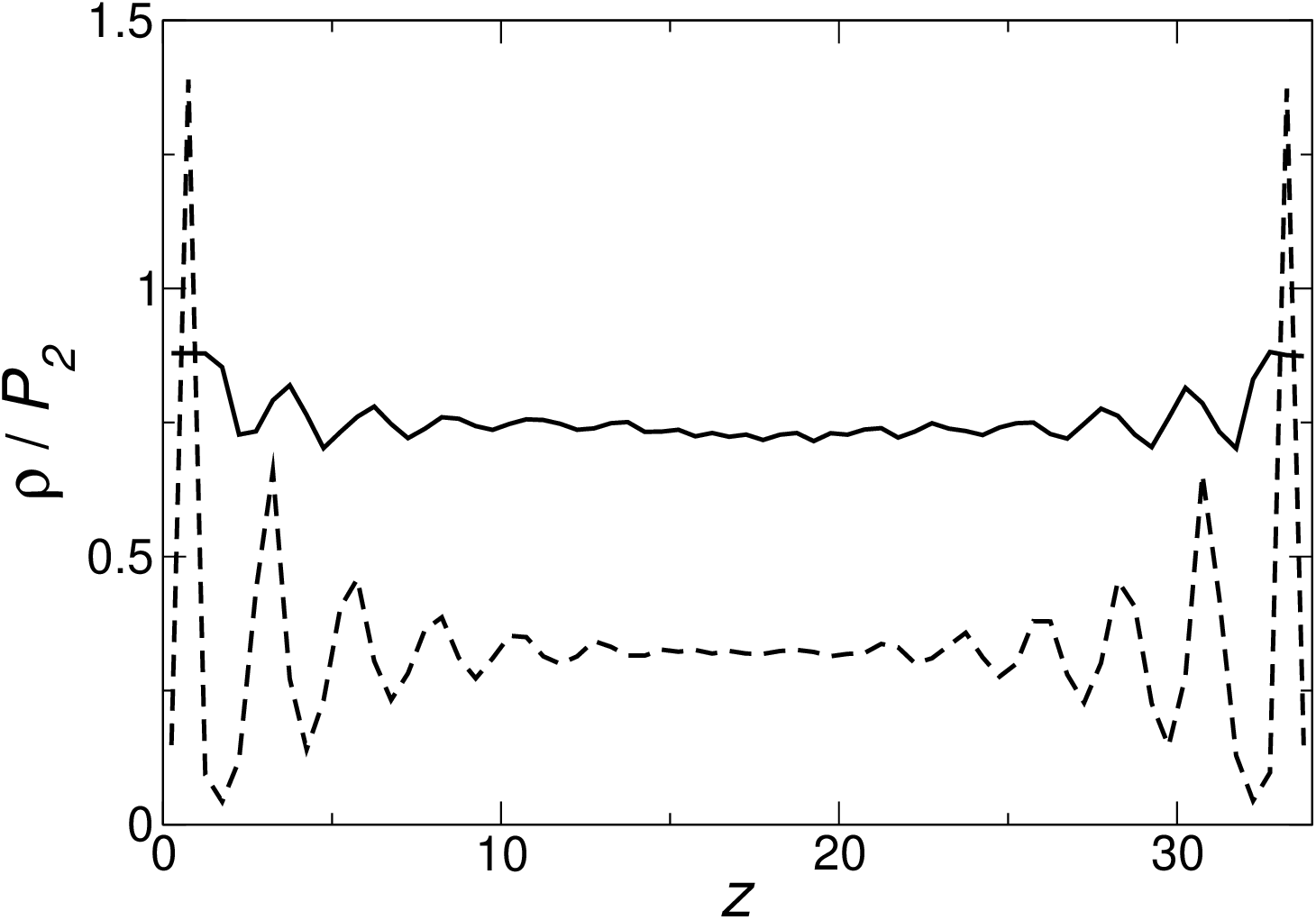}
  \caption{Density (dashed line) and $P_2$ (solid line) profiles for a confined system.}
  \label{fig:confined}
\end{figure}

Before proceeding with this system, we characterized the confined Gay-Berne
system without a colloid in independent simulations. The
separation between the walls was chosen to be $Z_{\text{wall}}=
34$. A second rank order parameter $P_2$ is defined as
\begin{equation}
  P_2 = \frac{1}{2}\langle 3(\unitvect{u}_i\unitvect{P})^2-1\rangle,
\end{equation}
where $\unitvect{P}$ is the average orientation of the system. 
Figure~\ref{fig:confined} shows the resulting profiles of
density and order parameter.
One can see that several layers are formed at the surface. In the
midsection between the walls both density and $P_2$ profiles are almost flat. 
In that midsection, a sphere is unlikely to ``feel'' the presence of a
substrate. On the other hand, the movement of the sphere in
the vicinity of the surface may be severely limited or dictated by the
layered structure of the liquid crystals.

\section{Results and Discussion}
The quantity of interest is the free energy profile for a colloidal
sphere in a confined nematic liquid crystal as a function of
distance between the center of mass of the sphere and the surface.
From the results shown in Figure~\ref{fig:confined} the fluid
structure is rich near the surface and bulk-like for $z>13$. Hence, we 
concentrate on the region $3\le z_{\text{sph}}\le 13$, where
$z_{\text{sph}}$ denotes the position
of the center of mass of a sphere and plays the role of a reaction
coordinate. The lower bound is the radius of the sphere. As a
first trial, $NVT$ MC simulations were performed with
$z_{\text{sph}}$ fixed at specific values. A plot
\begin{figure}

  \vspace{1cm}
  \includegraphics[width=0.7\linewidth]{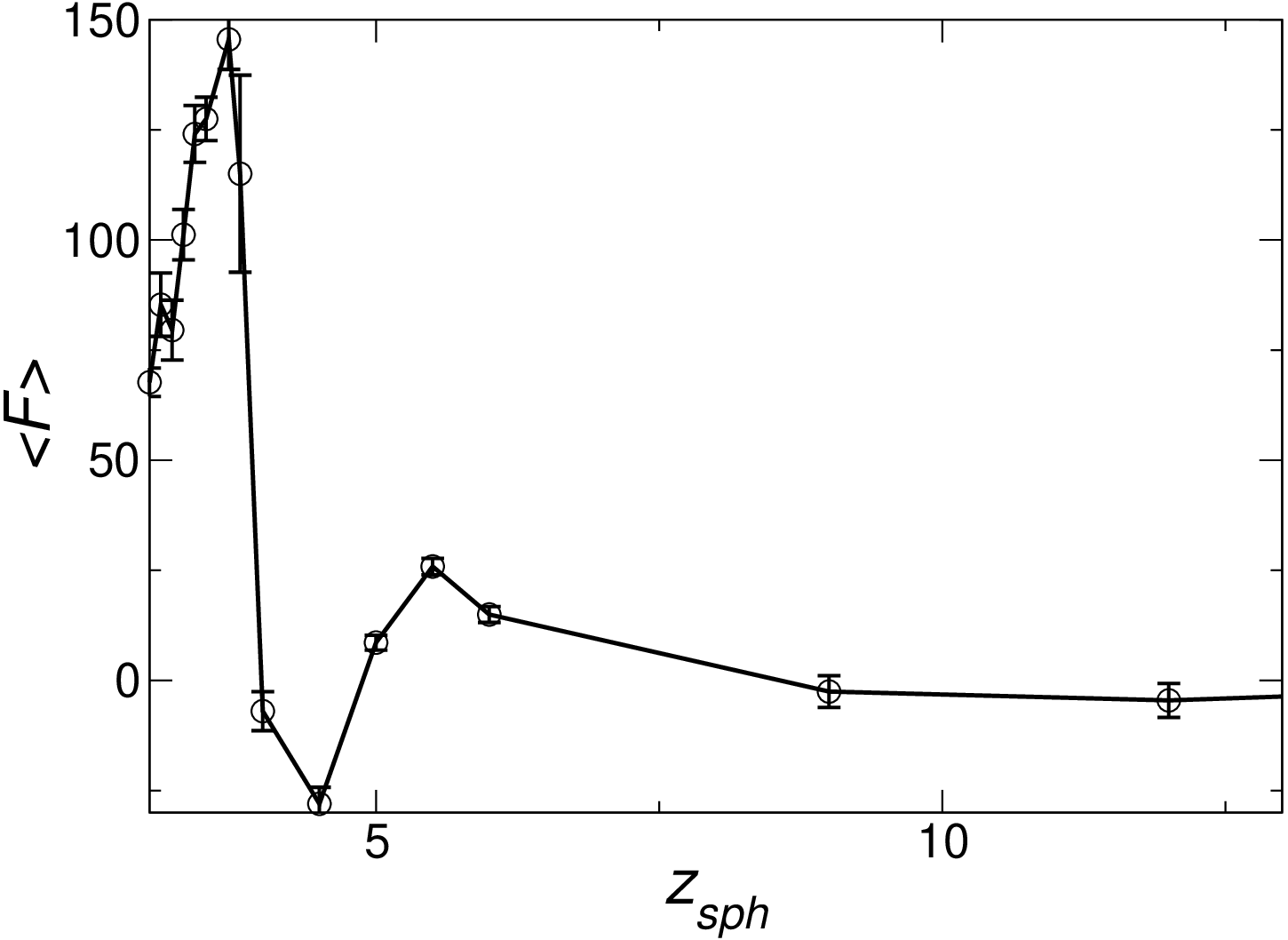}
  \caption{Mean force acting on a colloidal sphere from a  standard Monte 
  Carlo calculation.}
  \label{fig:meanforce}
\end{figure}
(Figure~\ref{fig:meanforce}) of the mean force acting on a sphere
from these calculations indicates that indeed the effective force
from the wall on a sphere is increasingly repulsive (positive) up
to $z_{\text{sph}} = 3.8$ and then decreases to zero around the
value of 4. The force then changes from repulsive to attractive
and back to repulsive and becomes seemingly constant and neutral
for $z_{\text{sph}}>9$. From the graph we can infer that at
$z_{\text{sph}}=4$ the system reaches a stable equilibrium. This
is confirmed by simulations of a freely moving sphere; from its original 
position of $z_{\text{sph}} = 3$, the sphere travels within a few MC
cycles to a position of $z_{\text{sph}} = 4$, where it undergoes only
slight oscillations. The structure of the liquid crystals near the surface
influences considerably the mean force on a sphere, but conventional
$NVT$ Monte Carlo simulations are unable to elucidate the details of these
interactions.

Next, we performed umbrella sampling simulations for $z_{\text{sph}}$
near the surface. In that region, the energy profile was found to vary
strongly. Highly stiff springs are necessary, which in turn result in
very narrow windows. To widen the range of the windows and improve
their overlap, we applied spring potentials of the form
$U_{\text{umb}} = A(z_{\text{sph}} - z_0)^4 + B(z_{\text{sph}} -
z_0)^2$, where $z_0$ is an equilibrium position for a spring and $A >
0$ and $B < 0$. Such potentials have a ``W'' shape and therefore
result in a slight dispersion of a sphere close to $z_0$. Depending on
the behavior of the system, at some points we had to use hybrid
springs that combined a stiff harmonic spring on one side of $z_0$ and
a dispersing potential on the other. As a result, 13 windows were
necessary to explore the narrow range of $z$ between $[3.0, 4.2]$. The
corresponding umbrella potentials are given in
Table~\ref{tab:umbrella}.
\setlength{\tabcolsep}{6pt}%
\begin{table}
  \def~{\kern0.5em}
  \begin{tabular}{rrrr}
  \multicolumn{1}{c}{No} & \multicolumn{1}{c}{$A$} & \multicolumn{1}{c}{$B$} & \multicolumn{1}{c}{$z_0$}\\
  \hline
  1     & 0 & 750 & 3.0~\\
  2\rlap{${}^*$} & 240\,000& $-1\,200$& 3.15\\
  2\rlap{${}'$} & 0 & 25\,000 & 3.15\\
  3\rlap{${}^*$} & 240\,000& $-1\,200$& 3.2~\\
  3\rlap{${}'$} & 0 & 3\,750 & 3.2~\\
  4 & 320\,000& $-1\,600$& 3.2~\\
  5 & 50\,000& $-1\,000$& 3.2~\\
  6 & 9\,877 & $-444$&3.2~\\
  7 & 3\,125 & $-250$ & 3.2~\\
  8 & 292\,969 & $-938$ & 3.5~\\
  9 & 15\,802 & $-178$ & 3.5~\\
 10 & 0 & 50\,000 & 3.7~\\
 11 & 2\,858 & $-185$ & 3.5~\\
 12\rlap{${}^*$} & 320\,000 & $-1600$ & 3.9~\\
 12\rlap{${}'$} & 0 & 25\,000 & 3.9~\\
 13 & 16 & $-18$ & 4.0~ \\
  \end{tabular}
  \caption{Values for the umbrella potentials $U_{\text{umb}}=
  A(z_{\text{sph}} - z_0)^4 + B(z_{\text{sph}} - z_0)^2$. Potentials marked
  with an asterisk are only applied to the left hand side of $z_0$, potentials
  marked with a prime are only applied to the right hand side of $z_0$. }
\label{tab:umbrella}
\end{table}

\begin{figure}
  \vspace{1.5cm}
  \includegraphics[width=0.7\linewidth]{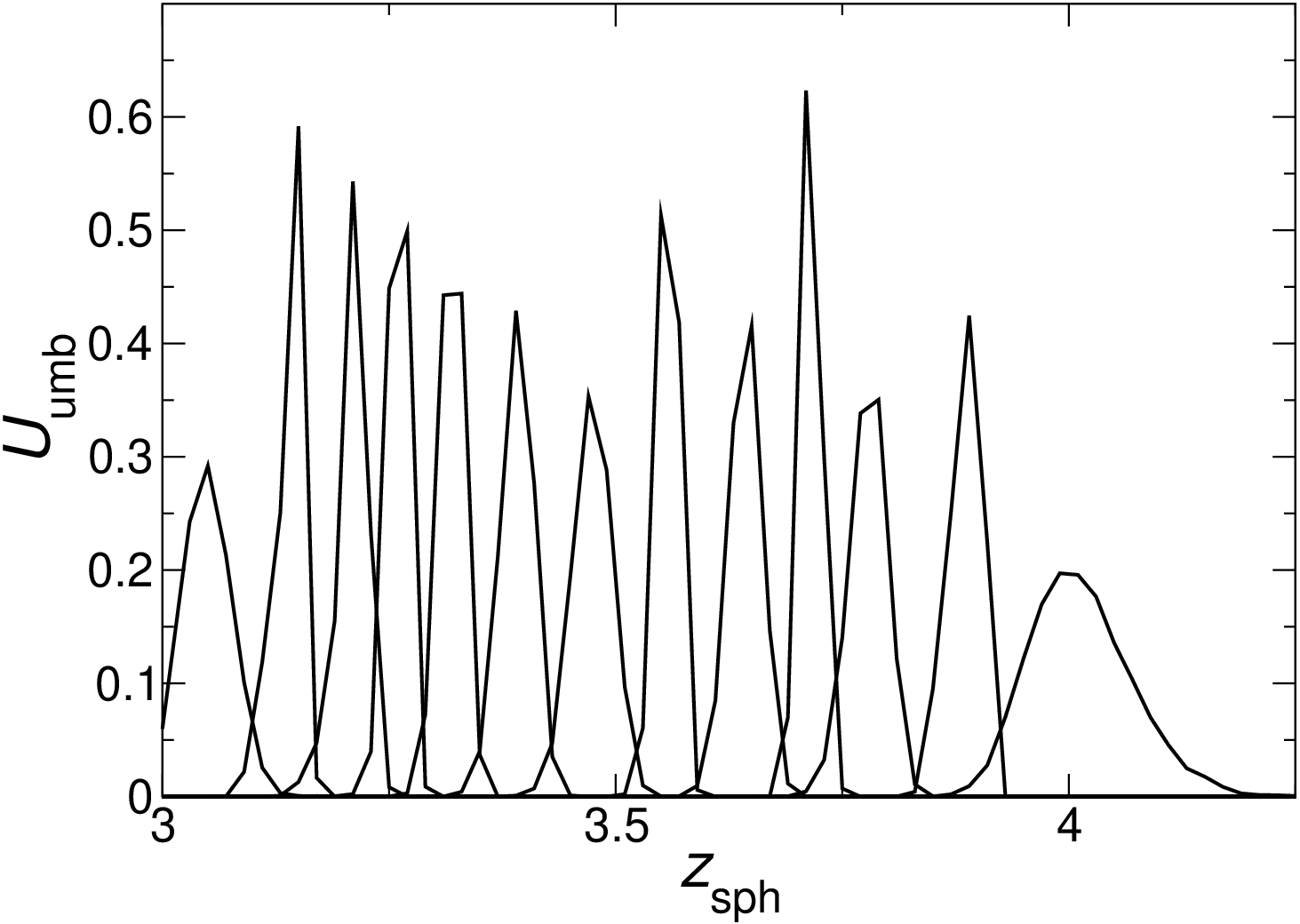}
  \caption{Energy histograms from umbrella sampling simulations for 13 
independently simulated umbrella potentials or windows.}
  \label{fig:umbrella}
\end{figure}
\begin{figure}
  \vspace{0.5cm}
  \includegraphics[width=0.4\linewidth]{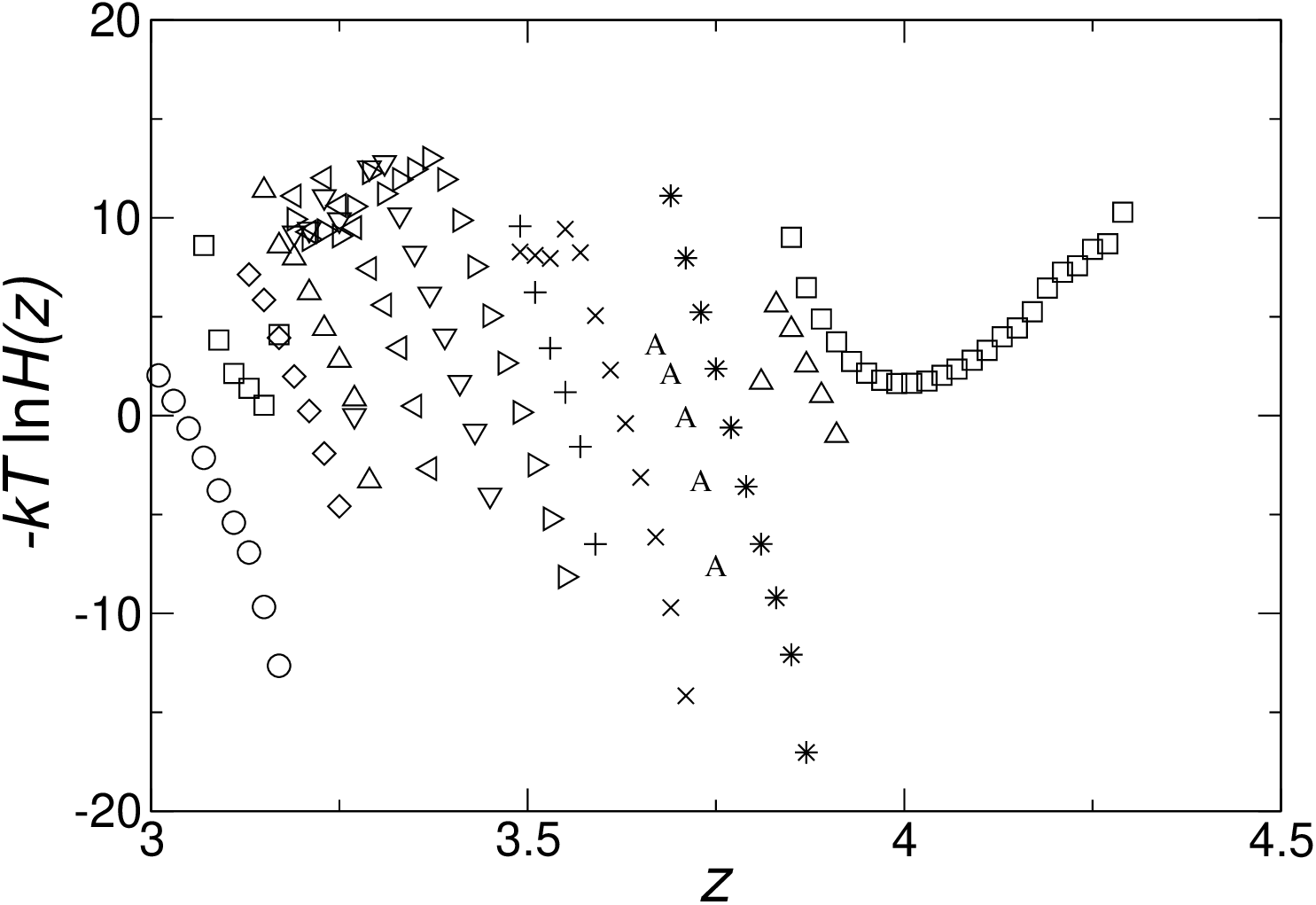}\hspace{0.5cm}
  \includegraphics[width=0.4\linewidth]{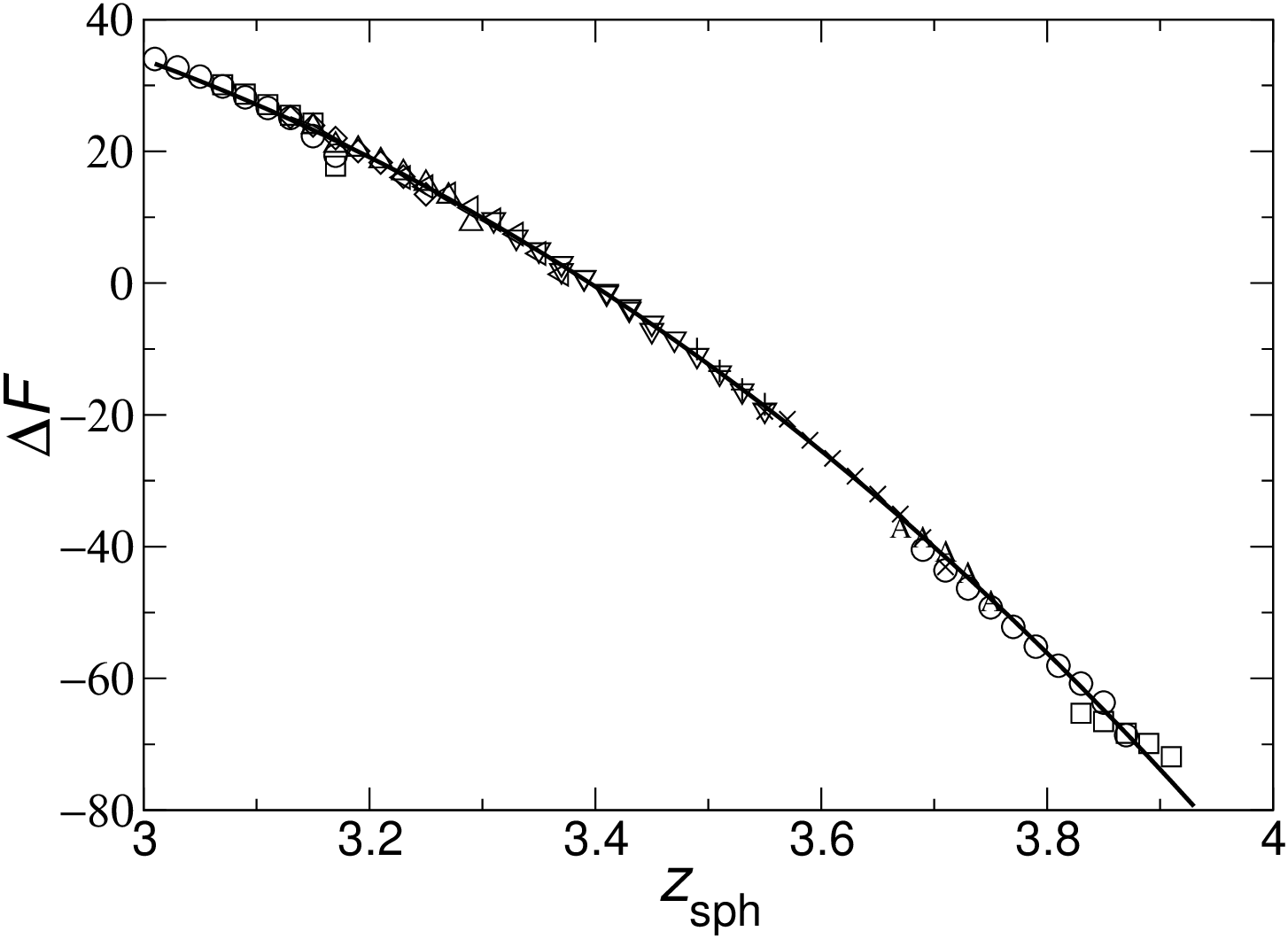}
  \caption{Left: original Boltzmann inverted histograms from umbrella sampling 
simulations. Right: The histograms of the 12 lower windows collapse onto a 
master plot to yield an effective relative free energy as a function of the 
distance of the colloid from one of the walls (see text).}
  \label{fig:umbrfit}
\end{figure}
\begin{figure}

  \vspace{0.5cm}
  \includegraphics[width=0.4\linewidth]{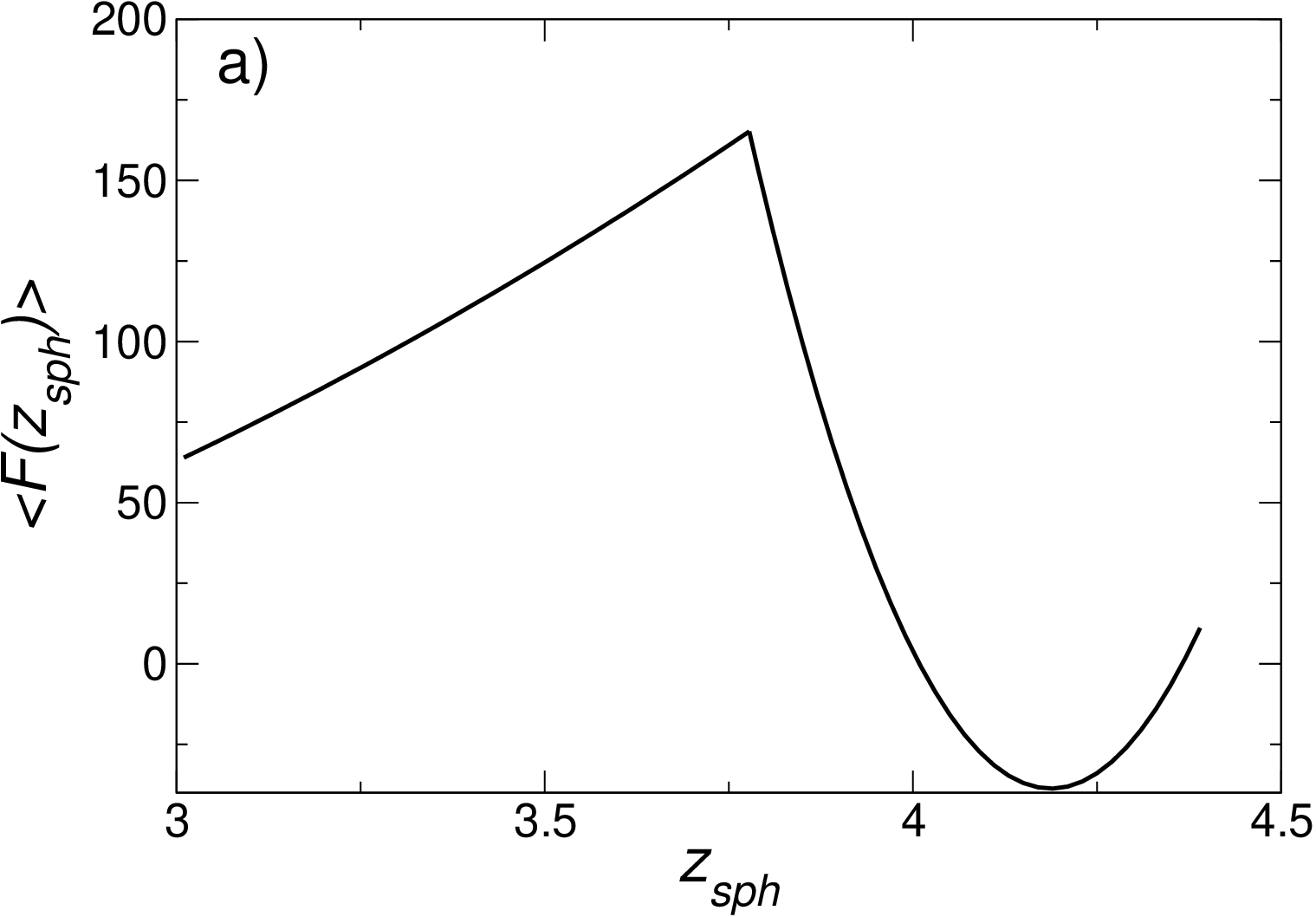}\hspace{0.5cm}
  \includegraphics[width=0.4\linewidth]{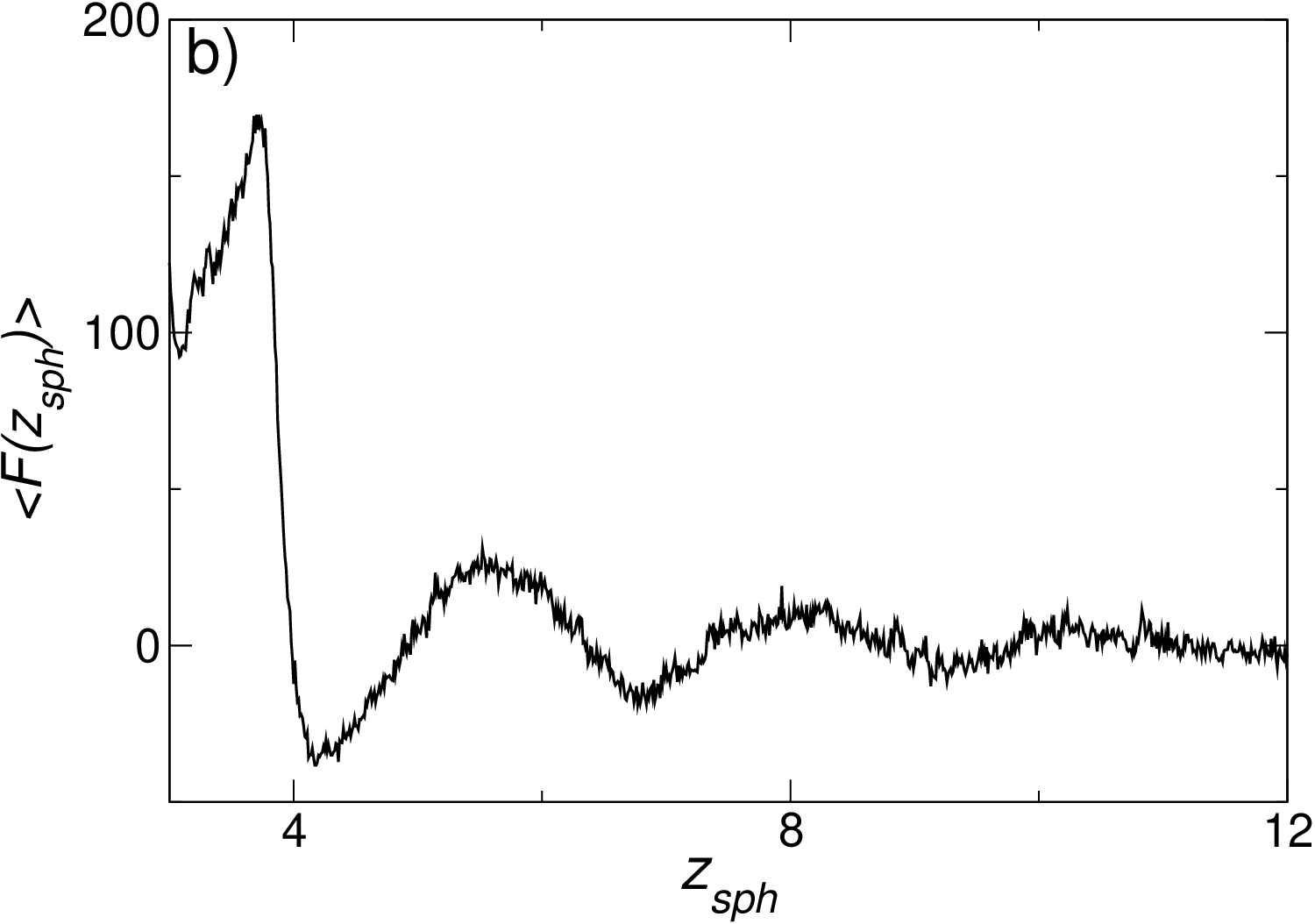}
  \caption{The mean force acting on the sphere calculated by the umbrella 
simulation technique (a) and from the novel technique (b)}
  \label{fig:forceumb}
\end{figure}

Figure~\ref{fig:umbrella} shows corresponding histograms collected
over a simulation of $3$--$5\times10^5$ MC cycles from the
initially equilibrated configurations. Not surprisingly, a
histogram centered at 4 is wide and Gaussian in appearance,
whereas the others are sharp and narrow. Histogram data
$g_i(z_{\text{sph}})$ are transformed by taking the natural
logarithm and subtracting the corresponding umbrella potentials.
Even though each data set is offset by a constant, there
appear to be at least two separate regions where different functions
must be fitted. We fit the first 12 sets and the last set
separately using third degree polynomials by a least squares procedure
modified to include constant offsets (see
Figure~\ref{fig:umbrfit}); the third degree was chosen as the
minimum required to have an inflection point, i.e. the derivative
of the force is zero.  The intersection of forces (negative
derivatives) is taken as the matching point; the curves of the potential of 
mean force are joined
at that point $(z_{\text{sph}} = 3.8, F(z_{\text{sph}} = 3.8) =
164)$ setting the PMF at $z_{\text{sph}} = 3$ to zero. The
resulting force profile (see Fig.~\ref{fig:forceumb}) is in
agreement with results of canonical simulations but reveals more
detail. 

\begin{figure}

  \vspace{2.5cm}
  \includegraphics[width=0.8\linewidth]{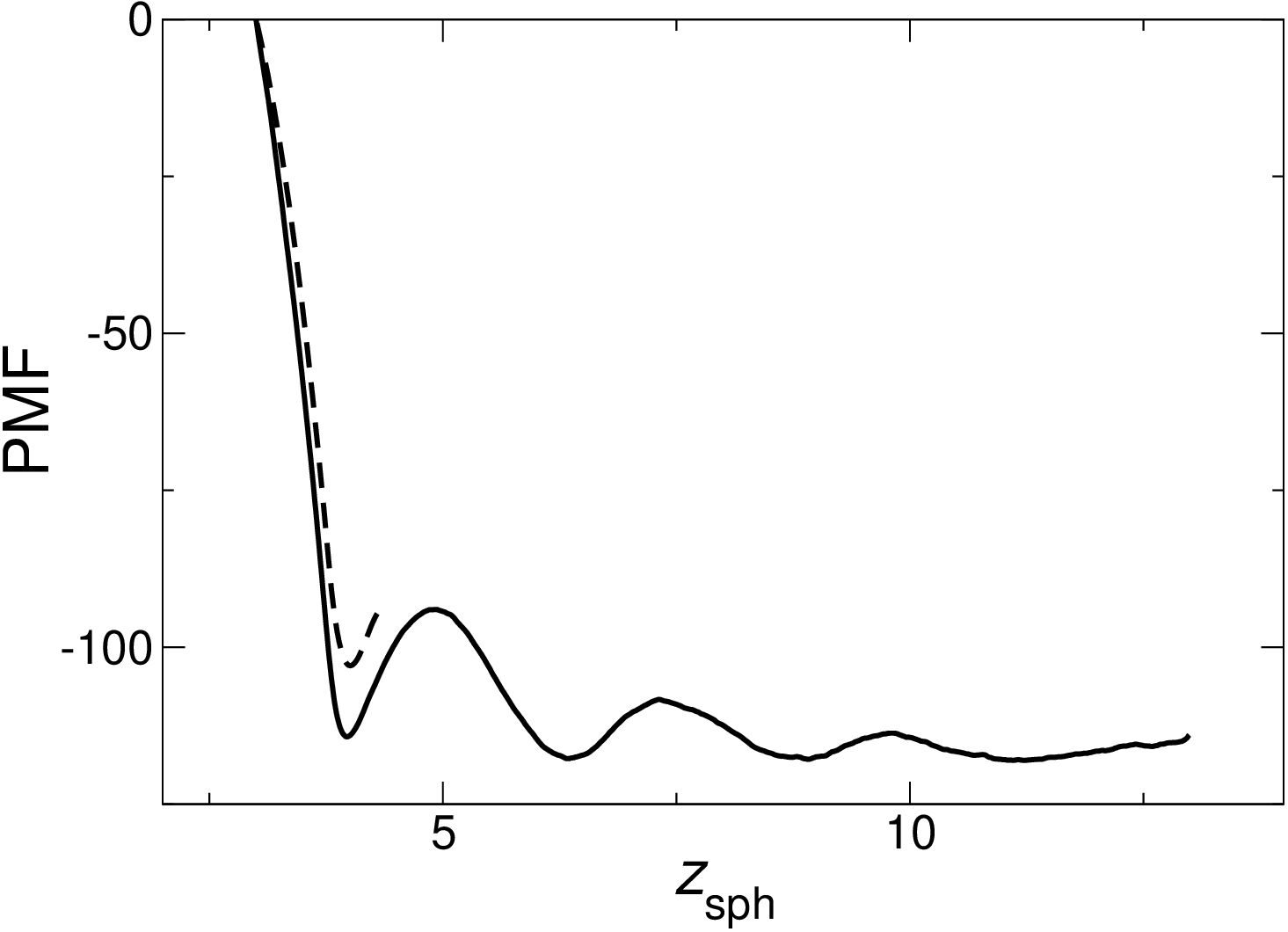}
  \caption{The potential of mean force between the the sphere and the surface 
calculated by the novel technique (solid line). For comparison the umbrella
calculated PMF is also shown (dashed line).}
  \label{fig:pmfcomp}
\end{figure}

Next, the application of the new method is demonstrated on the same
system. A state $m$ is defined by the position of the center of mass
of a sphere in the range $(m - 1) \Delta z < z_{\text{sph},m} \leq m
\Delta z$. The entire range of $[3, 13]$ is divided into nine
overlapping windows of width 2: the first window covers $3\le
z_{\text{sph}}<5$, the second window covers $4\le z_{\text{sph}}<6$,
etc. Each window consists of 200 states with $\Delta z=0.01$. A move
to a new state, i.e. a move of a sphere in the $z$ direction (its $x$
and $y$ coordinates are kept at zero throughout the entire simulation)
is attempted every Monte Carlo cycle; a cycle consists of a trial
displacement or rotation of all liquid crystal particles.  Initially,
all weighting factors $g_m$ and the convergence factor $f$ are given a
value of $e^{0.1}$. Simulations proceed for three DSMC cycles: that
is, $\ln f$ is halved twice. The resulting mean force and relative
weights are averaged between windows in the area of
overlap. Figure~\ref{fig:forceumb}b shows the results for the mean
force. The corresponding free energy profiles are shown in
Figure~\ref{fig:pmfcomp} (integrated mean force; PMF was arbitrarily
set to 0 at $z_{\text{sph}} = 3)$.  It is evident from the above plots
that the EDOS method provides improved sampling compared to simple
$NVT$ or umbrella potential simulations. Moreover, more details of
the interaction of the colloidal sphere with the liquid crystals and
the surface are visible (compare Figures~\ref{fig:meanforce} 
and~\ref{fig:forceumb}a). Note that with only 9 windows it is possible
to cover a substantially wider range than with 13 windows in umbrella
sampling.

It is apparent that the mean force on the sphere has different characteristics 
in various regions in proximity of the surface. Close to the surface, the force
is strongly repulsive and reaches its maximum ($F_{max} \approx 170$) at
$z_{\text{sph}} \approx 3.8$. The force then decreases rapidly, dropping to 
zero at $z_{\text{sph}}=4$ and eventually reaching a minimum
$(F_{min} \approx -35)$ at  $z_{\text{sph}} = 4.3$. In the interval 
$4 < z_{\text{sph}} < 11$ the force exhibits a damped oscillatory behavior 
with a wavelength of about 2.5--2.6. The force oscillations die out within 
about 2--3 of these wavelengths. At that point ($z_{\text{sph}} \approx 11$) 
the force becomes effectively zero.

\begin{figure}

 \includegraphics[width=\textwidth]{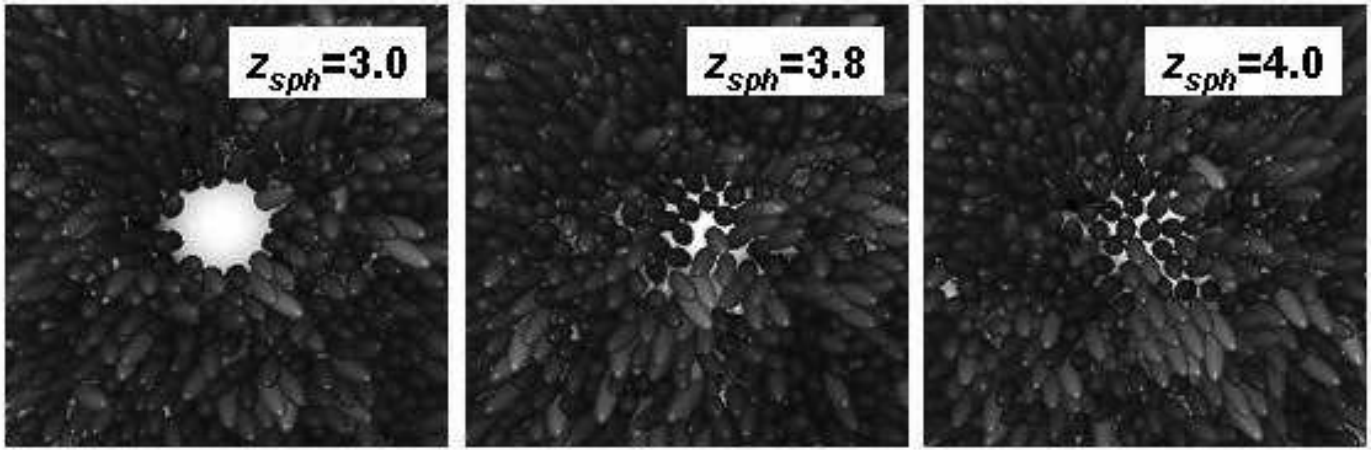}
 \caption{Views through the surface onto the sphere for 
$z_{\text{sph}}=3,3.8,4$ (left to right)} \label{fig:bottomview}
\end{figure}

Clearly, the layered structure of the liquid crystals has a strong
effect on the behavior of a suspended colloidal sphere. The effect
appears to be most pronounced in the immediate vicinity of the
surface, where the density and the order parameter are comparable to
those of a smectic or even solid layer.  Figure~\ref{fig:bottomview}
shows pictures of the colloid seen from below (the surface was removed
for clarity).  The sphere severely disrupts the first, dense layer of
liquid crystals; as the sphere moves away from the surface, the cavity
caused by it gradually fills up. At $z_{\text{sph}} = 3.8$ the layer
is almost completely restored. At $z_{\text{sph}} = 4$, the first
layer is fully repaired; this point corresponds to a stable state of
the system as the force on the sphere vanishes (see above).

\begin{figure}

  \vspace{3.5cm}
 \includegraphics[width=0.7\linewidth]{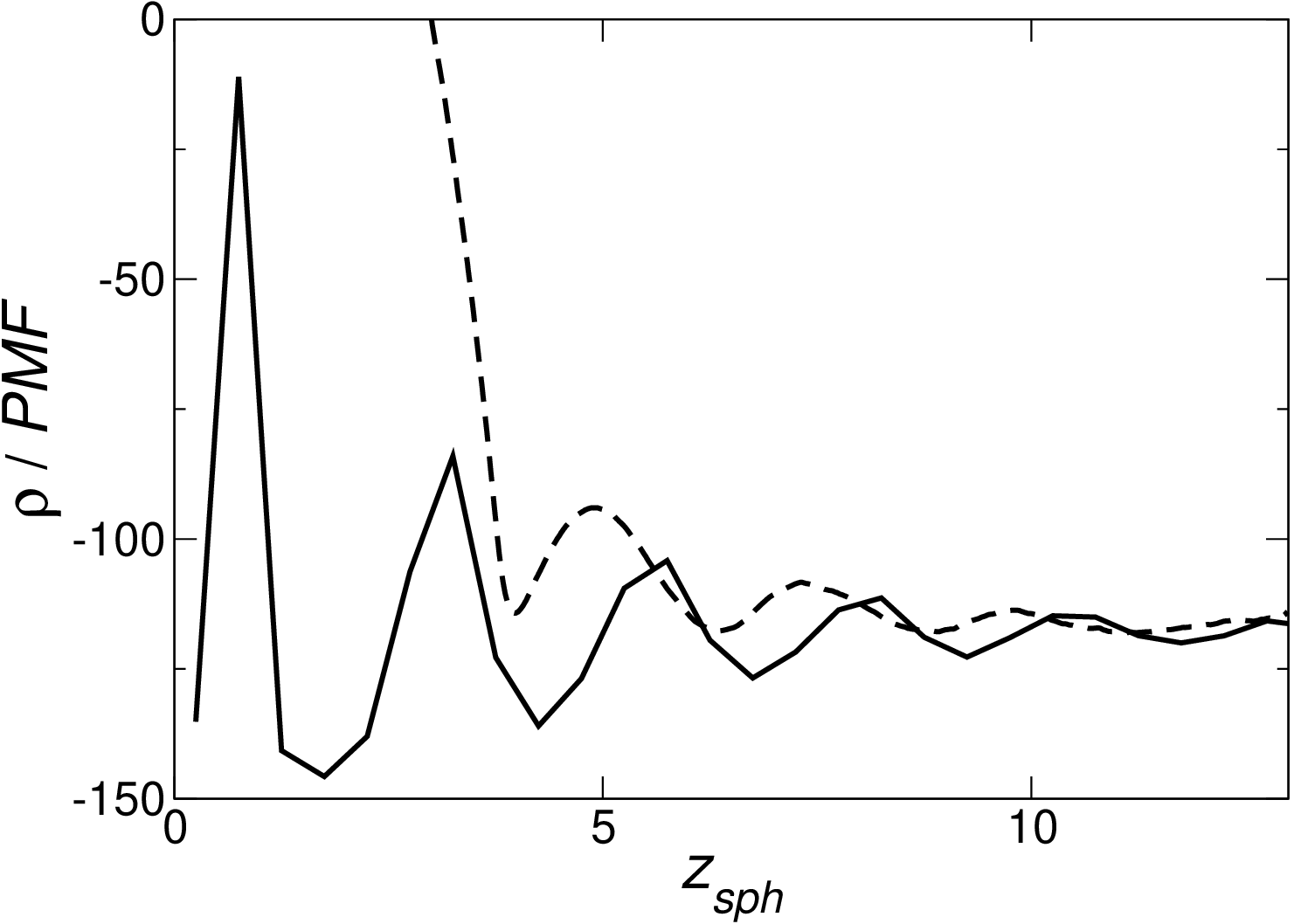}
 \caption{Density of the confined system without the sphere (solid line) 
overlaid with the potential of mean force gained by the novel method (dashed 
line). The density was scaled and shifted along $y$-axis arbitrarily.} 
\label{fig:overlay}
\end{figure}
As for subsequent undulations in the force profile, their wavelength 
approximately coincides with the thickness of the liquid crystalline layers
close to the surface. If we superimpose the density profile of the confined 
system without the sphere (see Figure~\ref{fig:overlay}) onto the PMF curve, 
the oscillations are almost in phase. The PMF curve is slightly shifted to
larger distances from the surface (offset in $z_{\text{sph}} \approx 0.25)$. 
The minima in the PMF practically match those of the density profile; our 
results indicate that the sphere prefers to reside between adjacent LC layers.
\section{Conclusion}
A new method is proposed in which a density of states Monte Carlo formalism is
used to self-consistently calculate the weights for an expanded 
ensemble technique. This method permits efficient calculation of the potential
of mean force of a complex system. Its application is illustrated in the 
context of a colloidal sphere dispersed in a confined nematic host.

The method allows one to determine with high resolution the mean force
on the colloid as a function of its separation from the surface; the
results agree qualitatively with less efficient methods such as
umbrella sampling and conventional canonical Monte Carlo. Clearly, the
layered structure of the liquid crystals near the surface profoundly
influences the mean force on the colloid. The snapshots of the system
at several distances indicate that the disruption of the first surface
layer is most likely responsible for the strong repulsive force very
close to the surface.  Once the ``hole'' caused by the sphere is
completely filled in, the force decreases to neutral at
$z_{\text{sph}}=4$, resulting in a preferred position for the colloid
reflected by a minimum in its free energy (potential of mean
force). At larger distances, both the force and the potential of mean
force profiles display undulations that are almost in phase with the
oscillations in the density profile of the confined system. It is
worth pointing out again that all intermolecular interactions in the
simulation are strictly repulsive. The attractive regions in the
potential of mean force between the colloid and the surface have
therefore purely entropic origins.  Additional studies are currently
under way to elucidate the nature of the interaction between the
surface and the colloid for different sphere sizes and different
ancoring conditions.
\section*{Acknowledgments}
This work was supported by the University of Wisconsin MRSEC for 
nanocomposites. RF also thanks the DFG (German Research Foundation) for 
financial support through the Emmy--Noether program.
\bibliography{standard}

\begin{thebibliography}{10}

\bibitem{gupta98}
V.~K. Gupta, J.~J. Skaife, T.~B. Dubrovsky, and N.~L. Abbott,
\newblock Science {\bf 279}, 2077 (1998).

\bibitem{vannelson02}
J.~A. {Van Nelson}, S.-R. Kim, and N.~L. Abbott,
\newblock Langmuir {\bf 18}, ASAP article (2002).

\bibitem{poulin97}
P.~Poulin, H.~Stark, T.~C. Lubensky, and D.~A. Weitz,
\newblock Science {\bf 275}, 1770 (1997).

\bibitem{poulin98}
P.~Poulin and D.~A. Weitz,
\newblock Phys Rev E {\bf 57}, 626 (1998).

\bibitem{gu00}
Y.~Gu and N.~L. Abbott,
\newblock Phys Rev Lett {\bf 85}, 4719 (2000).

\bibitem{billeter00}
J.~L. Billeter and R.~A. Pelcovits,
\newblock Phys Rev E {\bf 62}, 711 (2000).

\bibitem{andrienko01}
D.~Andrienko, G.~Germano, and M.~P. Allen,
\newblock Phys Rev E {\bf 63}, 041701 (2001).

\bibitem{roux94}
B.~Roux,
\newblock Comput Phys Commun {\bf 91}, 275 (1995).

\bibitem{engkvist96}
O.~Engkvist and G.~Karlstr{\"o}m,
\newblock Chem Phys {\bf 213}, 63 (1996).

\bibitem{vondele00}
J.~VandeVondele and U.~R{\"o}thlisberger,
\newblock J Chem Phys {\bf 113}, 4863 (2000).

\bibitem{straatsma92}
T.~P. Straatsma, M.~Zacharias, and J.~A. McCammon,
\newblock Chem Phys Lett {\bf 196}, 297 (1992).

\bibitem{sprik98}
M.~Sprik and G.~Ciccotti,
\newblock J. Chem. Phys. {\bf 109}, 7737 (1998).

\bibitem{wang01a}
F.~Wang and D.~P. Landau,
\newblock Phys. Rev. Lett. {\bf 86}, 2050 (2001).

\bibitem{deoliveira96}
P.~M.~C. {de Oliveira}, T.~J.~P. Penna, and H.~J. Herrmann,
\newblock Braz. J. Phys. {\bf 26}, 277 (1996).

\bibitem{chandler87}
D.~Chandler,
\newblock {\em Introduction to Modern Statistical Mechanics},
\newblock Oxford University Press, New York, 1987.

\bibitem{torrie74}
G.~M. Torrie and J.~P. Valleau,
\newblock Chem. Phys. Lett. {\bf 28}, 578 (1974).

\bibitem{beutler94}
T.~C. Beutler and W.~F. {Van Gunsteren},
\newblock J Chem Phys {\bf 100}, 1492 (1994).

\bibitem{berg92a}
B.~A. Berg and T.~Neuhaus,
\newblock Phys. Rev. Lett. {\bf 68}, 9 (1992).

\bibitem{ferrenberg88}
A.~M. Ferrenberg and R.~H. Swendsen,
\newblock Phys. Rev. Lett. {\bf 61}, 2635 (1988).

\bibitem{lyubartsev92}
A.~P. Lyubartsev, A.~A. Martinovski, S.~V. Shevkunov, and P.~N.
  Vorontsov-Velyaminov,
\newblock J. Chem. Phys. {\bf 96}, 1776 (1992).

\bibitem{escobedo95b}
F.~A. Escobedo and J.~J. {de Pablo},
\newblock J Chem Phys {\bf 103}, 2703 (1995).

\bibitem{wang01b}
F.~Wang and D.~P. Landau,
\newblock Phys Rev E {\bf 64}, 056101 (2001).

\bibitem{yan02}
Q.~Yan, R.~Faller, and J.~J. de~Pablo,
\newblock J Chem Phys {\bf 116}, 8745 (2002).

\bibitem{faller01sc}
R.~Faller and J.~J. de~Pablo,
\newblock Direct calculation of the density of states of a binary lennard-jones
  glass,
\newblock submitted.

\bibitem{rathore02}
N.~Rathore and J.~J. {de Pablo},
\newblock J Chem Phys {\bf 116}, 7225 (2002).

\bibitem{jain02}
T.~S. Jain and J.~J. {de Pablo},
\newblock J Chem Phys {\bf 116}, 7238 (2002).

\bibitem{kim02sb}
E.~B. Kim, R.~Faller, and J.~J. {de Pablo},
\newblock Estimation of the chemical potential of liquid crytal molecules by
  simulation,
\newblock in preparation.

\bibitem{faller02c}
R.~Faller, Q.~Yan, and J.~J. de~Pablo,
\newblock J Chem Phys {\bf 116}, 5419 (2002).

\bibitem{gay81}
J.~G. Gay and B.~J. Berne,
\newblock J Chem Phys {\bf 74}, 3316 (1981).

\end{thebibliography}
\bibliographystyle{aip}
\end{document}